\documentclass[prc,twocolumn,superscriptaddress,superscriptfootnote,nofootinbib]{revtex4}
\usepackage{amssymb}

\usepackage{amsmath,amsfonts,amssymb,epsfig,color,MnSymbol}

\begin{document}
\title{Shell-model representations of the microscopic version of the Bohr-Mottelson collective model}
\author{H. G. Ganev}
\affiliation{Joint Institute for Nuclear Research, Dubna, Russia}
\affiliation{Institute of Mechanics, Bulgarian Academy of Sciences,
Sofia, Bulgaria}

\setcounter{MaxMatrixCols}{10}

\begin{abstract}
The structure of the irreducible collective spaces of the group
$Sp(12,R)$, which many-particle nuclear states are classified
according to the chain $Sp(12,R) \supset U(6) \supset SO(6) \supset
SU_{pn}(3) \otimes SO(2) \supset SO(3)$ of the proton-neutron
symplectic model (PNSM), is considered in detail. This chain of the
PNSM was shown to correspond to a microscopic shell-model version of
the Bohr-Mottelson collective model. The construction of the
relevant shell-model representations of the $Sp(12,R)$ group along
this chain is considered for three nuclei with varying collective
properties and from different mass regions. It is shown that the
$SU_{pn}(3)$ basis states of the $Sp(12,R)$ representations are
always Pauli allowed for $\upsilon \geq \upsilon_{0}$, but organized
in a different way into different $SO(6)$ shells. This is in
contrast to the case of filling the levels of the standard
three-dimensional harmonic oscillator and using the plethysm
operation. Although the $SU_{pn}(3)$ multiplets with $\upsilon <
\upsilon_{0}$ are not all Pauli forbidden, it is safe to discard
them, as it was actually done in the practical applications.
\end{abstract}
\maketitle

PACS number(s): {21.60.Fw, 21.60.Ev}

\section{Introduction}

Different models of nuclear structure exist for describing a
particular set of experimental data or aspect of nuclear
excitations. These models can be roughly divided into two groups $-$
phenomenological and microscopic models. It is well known that a
characteristic feature that distinguishes between the two groups is
provided by the Pauli principle. The models are referred to as
microscopic if they fulfil the Pauli principle, which originates
from the fermionic nature of atomic nucleus. For the
phenomenological models the situation is opposite $-$ they do not
respect the Pauli principle, i.e. the composite fermion structure of
the nucleus is not taken into account. A well known example of a
microscopic model in nuclear physics is provided, for instance, by
the algebraic Elliott $SU(3)$ model of nuclear rotations
\cite{Elliott58}. A widely exploited phenomenological model of the
nuclear collective motion, which has conceptually influenced the
development of the other collective models of nuclear structure, is
presented by the Bohr-Mottelson (BM) collective model \cite{BM}. In
its standard formulation \cite{Bohr52,BM53}, the latter can not be
naturally related to the microscopic many-fermion nuclear theory. In
particular, it is not clear how the state vectors in the
Bohr-Mottelson model which characterize the quantized surface
vibrations and rotations of atomic nuclei can be identified with the
wave functions in the Hilbert space of $A$ nucleon antisymmetric
states. This is a common property of all phenomenological collective
models, which usually describe the nuclear collective motion in
terms of shape parameters or bosons of certain type.

It turns out that many phenomenological models of nuclear structure
can be given a microscopic foundation. For example, this is achieved
by considering the fermion composite substructure of the bosons
within the boson type models (see, e.g., \cite{IBM}). A more
powerful and elegant method to do this is provided by the algebraic
approach. The idea is to embed the desired phenomenological model
into the many-particle microscopic shell-model theory
\cite{stretched,Rowe96} by using the spectrum generating algebras
(SGA) and dynamical groups \cite{DGSGA}. It is well known that the
nuclear shell model (see, e.g., \cite{Heyde94}) provides such a
basic formal framework for understanding nuclei in terms of
interacting protons and neutrons. In this way the algebraic approach
appears as an unifying concept in the nuclear structure physics,
relating different models (irrespectively whether they are of
phenomenological or microscopic nature) by means of their algebraic
structures.

In algebraic models all model observables, such as Hamiltonian and
transition operators, are expressed in terms of the elements of a
Lie algebra of observables. In this regard, the problem of embedding
of a certain collective model is largely solved when once it is
recognized that both the collective model under consideration and
the shell model can be formulated as algebraic models with dynamical
groups. Thus, a certain collective model becomes a submodel of the
shell model if its dynamical group is expressed as a subgroup of a
dynamical group of the shell model (see, e.g.
\cite{stretched,Rowe96}). The full Lie algebra of observables of the
shell model is huge (strictly speaking, infinite), which is the
reason for making the shell model (with major-shell mixing) an
unsolvable problem and for seeking of its tractable approximations.
Fortunately, it has a subalgebra which is easier to manage; i.e.,
the Lie algebra of all one-body operators. The corresponding
dynamical group is then the group of one-body unitary
transformations. An example of a complete algebraic model that is a
submodel of the shell model is provided by the Elliott $SU(3)$ model
\cite{Elliott58} already mentioned above.

Then, in general, to give a certain collective model a microscopic
shell-model interpretation, the following three steps are required
within the framework of the algebraic approach: 1) algebraic
formulation of the collective model in terms of a Lie algebra of
observables; 2) seeking of a microscopic, many-particle realization
of this algebra in terms of all position and momentum coordinates of
the particles of the system; 3) construction of its shell-model
representations. Sometimes, it is necessary to adjust the considered
phenomenological model so that its algebraic structure becomes
compatible with the microscopic shell-model structure of the
nucleus. Such an example is provided by the embedding of the BM
model in the one-component shell-model theory. To make the BM model
microscopically realizable, one first needs to replace the shape
variables, which do not have a microscopic expression, by the
microscopic quadrupole moment operators. The latter together with
their time derivatives yield a new set of commutation relations,
defining the Lie algebra of the so called $CM(3)$ model of Weaver,
Biedenharn and Cusson \cite{cm3,NA4}. In this way one obtains a
microscopic many-particle realization of the BM model augmented by
the intrinsic vortex spin degrees of freedom, but which is not
compatible with the shell-model structure of the nucleus (i.e., only
the first two steps are performed). In order to make the $CM(3)$
model compatible with the fermion nature of the nucleus, its
dynamical group was extended to the non-compact symplectic group
$Sp(6,R)$ by including the many-particle kinetic energy operator to
its set of collective observables. In this way the $Sp(6,R)$ model
\cite{RR1}, sometimes called a microscopic collective model, is
obtained as a result of the embedding the BM model into the
one-component shell-model theory. It has well defined shell-model
representations, which are constructed by means of the
three-dimensional creation and annihilation operators of harmonic
oscillator quanta. The $Sp(6,R)$ model is a submodel of the nuclear
shell model, as should be according to the prescription described
above. Another example of embedding into the nuclear shell model is
presented by the phenomenological interacting vector boson model
\cite{IVBM}, in which the nuclear collective motion is described by
means of two types of vector bosons. This model possesses only one
shell-model representation for even-even nuclei $-$ namely, the
trivial scalar representation of its $Sp(12,R)$ dynamical group. A
microscopic foundation of this model was obtained by augmenting it
with an intrinsic microscopic many-particle $U(6)$ structure, which
already admits many nonscalar $Sp(12,R)$ representations compatible
with the fermion structure of the nucleus. This was achieved first
by expressing its SGA observables in terms of many-particle proton
and neutron position and momentum observables and then by
construction of the Pauli allowed shell-model representations (i.e.,
performing the second and third steps). As a result, a completely
new microscopic model of nuclear collective motion has appeared
which was referred to as a proton-neutron symplectic model (PNSM)
\cite{cdf,smpnsm}. At the same time, the PNSM generalizes the
microscopic one-component $Sp(6,R)$ model \cite{RR1} for the case of
the two-component proton-neutron nuclear systems, which becomes
evident by the embedding $Sp(6,R) \subset Sp(12,R)$. In this way the
PNSM has appeared as a simultaneous generalization of the
phenomenological interacting vector boson model and the  microscopic
$Sp(6,R)$ symplectic model.

Recently, the BM model was embedded \cite{microBM,mbm-cr} into the
two-component shell-model theory within the framework of the PNSM.
It was demonstrated that a microscopic shell-model version of the BM
model is defined by one of dynamical symmetry chains of the PNSM. It
is the purpose of the present paper to consider in detail the
shell-model irreducible representations of this new version and to
show that for $\upsilon \geq \upsilon_{0}$ they represent Pauli
allowed many-particle subspaces of the Hilbert space of the nucleus.

\section{The Proton-Neutron Symplectic Model}

The $Sp(12,R)$ SGA of the PNSM has many subalgebra chains, which can
be divided in two types $-$ the collective-model and shell-model
chains, respectively. For the shell-model purposes, the $Sp(12,R)$
SGA of the PNSM can be represented by its complexification
$Sp(12,R)=\{F_{ij}(\alpha,\beta),G_{ij}(\alpha,\beta),A_{ij}(\alpha,\beta)\}$.
In this realization, the $Sp(12,R)$ generators\cite{smpnsm}
\begin{align}
&F_{ij}(\alpha,\beta)=\sum_{s=1}^{m}b^{\dagger}_{i\alpha,s}b^{\dagger}_{j\beta,s},
\label{Fs} \\
&G_{ij}(\alpha,\beta)=\sum_{s=1}^{m}b_{i\alpha,s}b_{j\beta,s},
\label{Gs} \\
&A_{ij}(\alpha,\beta)=\frac{1}{2}\sum_{s=1}^{m}
(b^{\dagger}_{i\alpha,s}b_{j\beta,s}+b_{j\beta,s}b^{\dagger}_{i\alpha,s}).
\label{As}
\end{align}
are expressed as bilinear combinations of the standard creation and
annihilation operators of harmonic oscillator quanta
\begin{align}
&b^{\dagger}_{i\alpha,s}=
\sqrt{\frac{M_{\alpha}\omega}{2\hbar}}\Big(x_{is}(\alpha)
-\frac{i}{M_{\alpha}\omega}p_{is}(\alpha)\Big), \notag\\
&b_{i\alpha,s}=\sqrt{\frac{M_{\alpha}\omega}{2\hbar}}\Big(x_{is}(\alpha)
+\frac{i}{M_{\alpha}\omega}p_{is}(\alpha)\Big). \label{bos}
\end{align}
In the last expressions, $x_{is}(\alpha)$ and $p_{is}(\alpha)$
denote the coordinates and corresponding momenta of the
translationally-invariant relative Jacobi vectors of the
$m$-quasiparticle two-component nuclear system and $A$ is the number
of protons and neutrons. The range of indices is as follows: $i,j =
1,2,3$; $\alpha,\beta = p,n$ and $s = 1,\ldots,m=A-1$.

The microscopic shell-model version of the BM model is defined by
the following dynamical symmetry chain \cite{microBM,mbm-cr}:
\begin{align}
Sp(12,R) & \supset SU(1,1) \otimes SO(6) \notag\\
\langle\sigma\rangle \quad &\qquad\quad \lambda_{\upsilon}
\qquad\quad \upsilon \notag\\
\notag\\
&\supset U(1) \otimes SU_{pn}(3) \otimes SO(2) \supset SO(3),
\label{Sp2RxO6-DS} \\
&\qquad p \qquad \ (\lambda,\mu) \qquad\quad \nu \quad \ q \quad \ L
\notag
\end{align}
which represents a PNSM shell-model coupling scheme. The labels
under the different groups stand for their irreducible
representations. According to the chain (\ref{Sp2RxO6-DS}) the
combined monopole-quadrupole nuclear dynamics splits into radial and
orbital motions and the wave functions can be represented in the
form \cite{microBM}:
\begin{equation}
\Psi_{\lambda_{\upsilon}p;\upsilon\nu qLM}(r,\Omega_{5}) =
R^{\lambda_{\upsilon}}_{p}(r)Y^{\upsilon}_{\nu qLM}(\Omega_{5}).
\label{SU11xSO6wf}
\end{equation}
For more details concerning the structure of these function we refer
the readers to Ref.\cite{mbm-cr}.

The radial $SU(1,1)$ Lie algebra is generated by the shell-model
operators \cite{microBM}:
\begin{align}
S^{(\lambda_{\upsilon})}_{+} = \frac{1}{2}\sum_{\alpha}
F^{0}(\alpha,\alpha),
\label{Sp}\\
S^{(\lambda_{\upsilon})}_{-} = \frac{1}{2}\sum_{\alpha}
G^{0}(\alpha,\alpha),
\label{S-}\\
S^{(\lambda_{\upsilon})}_{0} = \frac{1}{2}\sum_{\alpha}
A^{0}(\alpha,\alpha), \label{S0}
\end{align}
which are obtained from (\ref{Fs})$-$(\ref{As}) by contraction with
respect to both indices $i$ and $\alpha$. The orbital motion group
$SO(6)$ can be expressed through the $U(6)$ generators
$A^{LM}(\alpha,\beta)$ (\ref{As}) in a standard way by taking their
antisymmetric combination \cite{microBM}:
\begin{equation}
\Lambda^{LM}(\alpha,\beta)=A^{LM}(\alpha,\beta)
-(-1)^{L}A^{LM}(\beta,\alpha).  \label{O6gen}
\end{equation}
The generators of different $SO(6)$ subgroups along the chain
(\ref{Sp2RxO6-DS}) are given by the following operators
\begin{align}
&\widetilde{q}^{2M}= \sqrt{3} i[A^{2M}(p,n)-A^{2M}(n,p)],
\label{q-SUpn3}
\\
&L^{1M}=\sqrt{2}[A^{1M}(p,p)+A^{1M}(n,n)], \label{Y-SUpn3}
\end{align}
and
\begin{equation}
M=-\sqrt{3}\Lambda^{0}(\alpha,\beta) =
-i\sqrt{3}[A^{0}(\alpha,\beta)-A^{0}(\beta,\alpha)], \label{O2gen}
\end{equation}
which generate the $SU_{pn}(3)$ and $SO(2)$ groups, respectively. As
can be seen from (\ref{q-SUpn3})$-$(\ref{Y-SUpn3}) and
(\ref{O2gen}), the two sets of operators are irreducible tensors of
different rank with respect to the group $SO(3)$. The two groups
$SU_{pn}(3)$ and $SO(2)$, therefore, are mutually complementary
\cite{MQ70} within the fully symmetric $SO(6)$ irreps $\upsilon
\equiv (\upsilon,0,0)_{6}$. The $SU_{pn}(3)$ irrep labels
$(\lambda,\mu)$ in this case are in one-to-one correspondence with
the $SO(6)$ and $SO(2)$ quantum numbers $\upsilon$ and $\nu$, given
by the following expression \cite{microBM}:
\begin{equation} (\upsilon )_{6}=\bigoplus_{\nu =\pm\upsilon
,\pm(\upsilon-2),...,0(\pm1)}(\lambda=\frac{\upsilon +\nu
}{2},\mu=\frac{\upsilon -\nu }{2})\otimes (\nu )_{2}.
\label{O6SUpn3}
\end{equation}
The reduction rules for $SU_{pn}(3) \supset SO(3)$ are given in
terms of a multiplicity index $q$ which distinguishes the same $L$
values in the $SU_{pn}(3)$ multiplet $(\lambda,\mu)$
\cite{Elliott58}:
\begin{eqnarray}
q &=&\min(\lambda,\mu),\min(\lambda,\mu)-2,...,0~(1)  \notag \\
L &=&\max(\lambda,\mu ),\max(\lambda,\mu)-2,...,0~(1); \ q=0 \qquad
\label{su3o3}
\\
L &=&q,q+1,...,q+\max(\lambda,\mu); \ q\neq 0.  \notag
\end{eqnarray}

For our present purposes, however, it is more convenient to use the
equivalent \cite{microBM,sp2rxso6} dynamical chain
\begin{equation}
Sp(12,R) \supset U(6) \supset SO(6) \supset SU_{pn}(3) \otimes SO(2)
\supset SO(3) \label{O6-DS}
\end{equation}
to classify the many-particle shell-model states of the nucleus. The
branching rules for the reduction $U(6) \supset SO(6)$ in the case
of fully symmetric representations $[E]_{6}$ of $U(6)$ are given by
\cite{Van71}:
\begin{equation}
[E]_{6}=\bigoplus_{\upsilon =E,E-2,...,0(1)}(\upsilon
,0,0)_{6}=\bigoplus_{i=0}^{\langle\frac{E}{2}\rangle}(E-2i)_{6},
\label{U6O6}
\end{equation}
where $\langle E/2\rangle = E/2$ if $E$ is even and $(E-1)/2$ if $E$
is odd. From the latter expression we see that only fully symmetric
$(\upsilon ,0,0)_{6}\equiv (\upsilon )_{6}$ irreps of $SO(6)$
appear. For non-symmetric $U(6)$ irreducible representations one can
use, e.g., the SCHUR computer program \cite{SCHUR} to obtain the
corresponding $SO(6)$ subrepresentations.

\section{Shell-model representations}

The symplectic basis for an irreducible representation $\langle
\sigma \rangle \equiv \langle
\sigma_{1}+\frac{m}{2},\ldots,\sigma_{6}+\frac{m}{2} \rangle$ of the
group $Sp(12,R)$ is constructed by acting on the $Sp(12,R)$
lowest-weight state $|\sigma \rangle$ by the symplectic raising
operators (\ref{Fs}). This can be symbolically represented in the
following form \cite{smpnsm}:
\begin{equation}
|\Psi(\sigma n \rho E \eta) \rangle= [P^{(n)}(F) \times |\sigma
\rangle]^{\rho E}_{\eta} , \label{sbasis}
\end{equation}
where $P^{(n)}(F) = [F\times \ldots \times F]^{(n)}$ and $n =
[n_{1},\ldots,n_{6}]$ is a partition with even integer parts. $E =
[E_{1},\ldots,E_{6}]$ indicates the $U(6)$ quantum numbers of the
coupled state, $\eta$ labels a basis of states for the coupled
$U(6)$ irrep $E$, and $\rho$ is a multiplicity index. In this way we
obtain a basis of $Sp(12,R)$ states that reduces the subgroup chain
$Sp(12,R) \supset U(6)$. In addition, in our practical applications
we usually restrict the model space only to the fully symmetric
$U(6)$ irreps $E = [E_{1}\equiv E, 0, \ldots,0] \equiv [E]_{6}$.

The symplectic basis states are further classified by the remaining
groups in the chain (\ref{O6-DS}). This means that the symplectic
states are characterized by their irreducible representations, i.e.
we fix the basis index $\eta=\upsilon\nu qLM$ in Eq.(\ref{sbasis}).
But using the relation (\ref{O6SUpn3}), one alternatively obtains
for the basis index $\eta=\upsilon(\lambda,\mu) qLM$. The latter
choice is more convenient for the analysis of the $SU(3)$ content of
the shell-model representations of $Sp(12,R)$. We note also that the
lowest-weight state of $Sp(12,R)$ is simultaneously a highest-weight
state for the $U(6)$ irreducible representation $\sigma \equiv
[\sigma_{1}, \ldots, \sigma_{6}]$. Such a $Sp(12,R)$ lowest-weight
but $U(6)$ highest-weight state is sometimes referred to as a
lowest-grade $U(6)$ state. For this intrinsic $U(6)$ structure we
will simply use the term symplectic bandhead or $Sp(12,R)$ bandhead.

To understand better the type and the structure of shell-model
representations of the $Sp(12,R)$ basis states that are classified
either by the dynamical chain (\ref{Sp2RxO6-DS}) or (\ref{O6-DS}),
we will consider the relevant representations for three nuclei with
varying collective properties and belonging to different mass
regions. First consider the relevant $Sp(12,R)$ shell-model
irreducible representation for the light nucleus $^{20}$Ne.

\subsection{Shell-model representation of $^{20}$Ne}

It is well known that possible $SU(3)$ states in the nuclear shell
model are obtained by taking all possible distributions of protons
and neutrons within the considered valence shells. The set of Pauli
allowed states within a given three-dimensional oscillator shell
$\mathcal{N}$ can be obtained by the so-called plethysm operation,
according to which the set of the $SU(3)$ shell-model states are
defined by the reduction chain $U(d) \supset SU(3)$, where $d
=\frac{1}{2}(\mathcal{N}+1)(\mathcal{N}+2)$ for each nuclear shell
$\mathcal{N}$. Computer codes \cite{UNtoSU3a,UNtoSU3b} exist for the
evaluation of the $SU(3)$ irreps contained in $U(d)$. For the case
of two-component nuclear system, one should first consider $U_{p}(d)
\supset SU_{p}(3)$ and $U_{n}(d) \supset SU_{n}(3)$ ($\alpha=p,n$)
with the consequent coupling of the proton and neutron
$SU_{\alpha}(3)$ multiplets, i.e., $(\lambda_{p},\mu_{p})\otimes
(\lambda_{n},\mu_{n})$, to the $SU(3)$ irreducible representation
$(\lambda,\mu)$ of the combined proton-neutron nuclear system.
Generally, we have many possible proton-neutron $SU(3)$ multiplets,
i.e. $(\lambda_{p},\mu_{p})\otimes (\lambda_{n},\mu_{n})=
\sum(\lambda,\mu)$.

Filling pairwise the levels of three-dimensional harmonic oscillator
by protons and neutrons separately at the experimentally observed
quadrupole deformation, staring from bottom, we obtain completely
filled $s$ and $p$ shells, plus two protons and two neutrons in the
$sd$ shell. That is, we obtain the same many-particle configuration
$(0)^{2}(1)^{6}(2)^{2}$ for the proton and neutron subsystem. Then,
using the codes \cite{UNtoSU3a,UNtoSU3b}, for $^{20}$Ne one readily
obtains the following $SU(3)$ irreducible representations for the
proton (neutron) subsystem: $(4,0)$ and $(0,2)$. The Pauli allowed
$SU(3)$ multiplets for the combined proton-neutron nuclear system
are obtained by the direct products of these two irreps, i.e.: a)
$(4,0)\otimes (4,0) = (8,0), (6,1), (4,2), (2,3), (0,4)$; b)
$(4,0)\otimes (0,2) = (4,2), (3,1), (2,0)$; and c) $(0,2)\otimes
(0,2) = (0,4), (1,2), (2,0)$. Each one of these $SU(3)$ multiplets,
for example, can serve as an intrinsic $SU(3)$ structure for the
construction of an $Sp(6,R)$ shell-model representation. The
many-particle Hilbert space for $^{20}$Ne therefore can be
represented as a direct sum of different $Sp(6,R)$ shell-model
irreducible representations, including $-$ beyond the 0p-0h
representations built on the $sd$ valence shell $SU(3)$ multiplets
just obtained $-$ also the excited $Sp(6,R)$ representations by
taking all possible distributions of the protons and neutrons over
the higher major shells. Usually, the leading $SU(3)$ representation
is used, which is obtained by coupling the leading, i.e. most
deformed, proton and neutron representations. Thus, for $^{20}$Ne,
one obtains the leading proton-neutron $(8,0)$ multiplet. The
irreducible collective space within the $Sp(6,R)$ model, built upon
this $(8,0)$ multiplet, is given in Table \ref{Ne20Sp6RIR} as an
example.

\begin{table}[h!]
\caption{Irreducible collective space 0p-0h $(8,0)$ of $Sp(6,R)$,
relevant to $^{20}$Ne.} \label{Ne20Sp6RIR}
\smallskip\centering\small\addtolength{\tabcolsep}{0.pt}
\begin{tabular}{lllll}
\hline &  & $\quad \cdots \qquad\qquad\qquad\qquad\qquad\quad \  \cdots $ &  &  \\
\hline
&  & $%
\begin{tabular}{l}
$N_{0}+4\quad (12,0),(10,1),2(8,2),(6,3),(7,1),(4,4),(6,0)$
\end{tabular}%
$ &  &  \\ \hline
&  & $%
\begin{tabular}{l}
$N_{0}+2 \qquad\qquad\qquad (10,0),(8,1),(6,2)$
\end{tabular}%
$ &  &  \\ \hline &  & $\ \ N_{0}\qquad\qquad\qquad\qquad\qquad \ \
(8,0)$ & &
\\ \hline
\end{tabular}%
\end{table}

Alternatively, one can use the supermultiplet spin-isospin scheme to
obtain the Pauli allowed $SU(3)$ states. Thus, filling each level by
four nucleons, one obtains the many particle configuration:
$(0)^{4}(1)^{12}(2)^{4}$. For 4 nucleons in the $\mathcal{N}=2$ $sd$
shell, the codes \cite{UNtoSU3a,UNtoSU3b} produce: $(8,0), (4,2),
(0,4), (2,0)$. We see that the odd $SU(3)$ irreps obtained in the
proton-neutron scheme are now missing. The even $SU(3)$ irreps are
the same.

The relevant irreducible collective space for $^{20}$Ne, spanned by
the $Sp(12,R)$ irreducible representation 0p-0h $[12]_{6}$ (or using
an equivalent notation, $\langle \sigma\rangle = \langle 10+19/2,
2+19/2, \ldots, 2+19/2\rangle$) that is restricted only to the fully
symmetric $U(6)$ irreps and which basis states are classified by the
chain (\ref{O6-DS}), is given in Table \ref{Ne20-Sp12RIR}. This
$Sp(12,R)$ representation is defined by the intrinsic $U(6)$
structure $[10,2,2,2,2,2]_{6} \equiv [8]_{6}$, which in turn is
fixed by the leading $SU(3)$ irrep $(8,0)$. From the figure the
structure of the symplectic basis become evident. Some points are of
importance at this place. First, the collective potential that can
be expressed along the chain (\ref{O6-DS}) as a function of the
second- and third Casimir operators of $SU_{pn}(3)$ will organize
the space of $SU(3)$ irreps according to their deformation. That is,
the lowest in energy will be the $SU_{pn}(3)$ multiplet $(8,0)$ from
the maximal seniority $SO(6)$ irrep $\upsilon_{0} = 8$ of the
symplectic bandhead. We note that in the absence of the third-order
$SU_{pn}(3)$ Casimir operator that distinguishes between the prolate
and oblate shapes, the same energy will be obtained for the
conjugate multiplet $(0,8)$ from the $SO(6)$ irrep $\upsilon_{0} =
8$. The other $SU_{pn}(3)$ multiplets belonging to the $SO(6)$ irrep
$\upsilon_{0} = 8$ will be higher in energy, followed by the
$SU_{pn}(3)$ multiplets for other $SO(6)$ irreps with $\upsilon <
\upsilon_{0}$ belonging to the lowest-grade $U(6)$ irrep $[8]_{6}$
characterized also by $N_{0}$. For the other $U(6)$ shells the
situation will be similar. Note that the different major shells are
separated by the harmonic oscillator energy $\hbar\omega =
41A^{-1/3}$ and the $n$-th excited shell will have an energy $n
\hbar\omega$.

Second, the horizontal set of $SU(3)$ irreducible representations
of, e.g., the row defined by $N_{0}$ at first sight look differently
compared to that obtained by the plethysm operation via the
reduction $U(d) \supset SU(3)$ \cite{UNtoSU3a,UNtoSU3b} and given
above. That is, the overlap of the two sets of $SU(3)$ irreps looks
partial. This is because the many-particle configurations in the
PNSM are classified by the basis states of the six-dimensional
harmonic oscillator rather than the standard three-dimensional one.
But the $SU(3)$ states contained in the $U(6)$ group structure can
be organized in different ways since different choices for the group
$G$ in the reduction $U(6) \supset G \supset SU(3)$ are possible.
Then each shell in the present approach is determined by the
corresponding $U(6)$ representation, which in turn contains
different seniority $SO(6)$ irreducible representations $\upsilon$
or subshells (see Table \ref{Ne20-Sp12RIR}). Consider first the
$SU(3)$ irreps belonging to the maximal seniority $SO(6)$ irrep
$\upsilon_{0} = 8$ of the $Sp(12,R)$ bandhead structure $N_{0}$,
which is of particular interest in the practical application of the
microscopic shell-model version of the BM model. We will show now
that the horizontal set of the remaining $SU(3)$ irreps which are
placed to the right from the axially-symmetric multiplet
$(\lambda=8,0)$, the latter being in the most left position,
actually represent many-particle-many-hole (mp-mh) excitations of
the nuclear system.

\begin{widetext}
\begin{center}
\begin{table}[h!]
\caption{Irreducible collective space 0p-0h $[8]_{6}$ of $Sp(12,R)$,
relevant to $^{20}$Ne, which $SU_{pn}(3)$ basis states are
classified according to the chain (\ref{O6-DS}).}   \label{Ne20-Sp12RIR} % title of Table
\end{table}
\vspace{-30pt}
%\centering
\smallskip\centering\small\addtolength{\tabcolsep}{4.pt}
%\begin{tabular}{c c c c}            % centered columns (4 columns)
\begin{equation*}
\smallskip \centering{\small \addtolength{\tabcolsep}{0.8pt}
\begin{tabular}{||l|l|l||}
\hline\hline $\ \ \ N$ & $\upsilon \backslash \nu $ &
\begin{tabular}{lllllllllllll}
$\cdots $ &  $\ 10$ &  $\ \ \ \ \ \ 8$ &  $\ \ \ \ \ 6$ &  $\ \ \ \
\ 4$ &  $\ \ \ \ \ 2$ &  $\ \ \ \ \ \ 0$ &  $\ \ \ \ -2$ &  $\ \ \ \
\ -4$ & \ \  $\ -6$ &  $\
\ \ -8$ &  $\ \ \ -10$ & $\ \cdots $%
\end{tabular}
\\ \hline\hline
$\ \ \ \ \vdots $ & $\ \ \vdots $ & $%
\begin{tabular}{lllllllllllll}
$\ddots $ &  $\ \ \ \vdots $ &  \ \ \ \ \ \ \ $\vdots $ &  \ \ \ \ \
\ $\vdots $ & \ \ \ \ \ \ $\vdots $ &  \ \ \ \ \ \ $\vdots $ &  \ \
\ \ \ \ \ $\vdots $ &  \ \ \ \ \ \ $\vdots $ &  \ \ \ \ \ \ $\vdots
$ & \ \ \ \ \ \ $\vdots $ &  \ \ \ \
\ \ \ $\vdots $ &  \ \ \ \ \ \ $\vdots $ &  \ \ \ $\udots $%
\end{tabular}%
$ \\ \hline $N_{0}+2$ &
\begin{tabular}{l}
$10$ \\
$8$ \\
$6$ \\
$4$ \\
$2$ \\
$0$%
\end{tabular}
&
\begin{tabular}{ll}
& $(10,0)$ \\
&  \\
&  \\
&  \\
&  \\
. & .%
\end{tabular}%
\begin{tabular}{l}
$(9,1)$ \\
$(8,0)$ \\
\\
\\
\\
.%
\end{tabular}%
\begin{tabular}{l}
$(8,2)$ \\
$(7,1)$ \\
$(6,0)$ \\
\\
\\
.%
\end{tabular}%
\begin{tabular}{l}
$(7,3)$ \\
$(6,2)$ \\
$(5,1)$ \\
$(4,0)$ \\
\\
.%
\end{tabular}%
\begin{tabular}{l}
$(6,4)$ \\
$(5,3)$ \\
$(4,2)$ \\
$(3,1)$ \\
$(2,0)$ \\
.%
\end{tabular}%
\begin{tabular}{l}
$(5,5)$ \\
$(4,4)$ \\
$(3,3)$ \\
$(2,2)$ \\
$(1,1)$ \\
$(0,0)$%
\end{tabular}%
\begin{tabular}{l}
$(4,6)$ \\
$(3,5)$ \\
$(2,4)$ \\
$(1,3)$ \\
$(0,2)$ \\
.%
\end{tabular}%
\begin{tabular}{l}
$(3,7)$ \\
$(2,6)$ \\
$(1,5)$ \\
$(0,4)$ \\
\\
.%
\end{tabular}%
\begin{tabular}{l}
$(2,8)$ \\
$(1,7)$ \\
$(0,6)$ \\
\\
\\
.%
\end{tabular}%
\begin{tabular}{l}
$(1,9)$ \\
$(0,8)$ \\
\\
\\
\\
.%
\end{tabular}%
\begin{tabular}{l}
$(0,10)$ \\
\\
\\
\\
\\
.%
\end{tabular}
\\ \hline
$\ \ N_{0}$ &
\begin{tabular}{l}
$8$ \\
$6$ \\
$4$ \\
$2$ \\
$0$%
\end{tabular}
& \ \ \ \ \ \ \ \ \ \ \ \ \ \ \ \ \ \
\begin{tabular}{l}
$(8,0)$ \\
\\
\\
\\
.%
\end{tabular}%
\begin{tabular}{l}
$(7,1)$ \\
$(6,0)$ \\
\\
\\
.%
\end{tabular}%
\begin{tabular}{l}
$(6,2)$ \\
$(5,1)$ \\
$(4,0)$ \\
\\
.%
\end{tabular}%
\begin{tabular}{l}
$(5,3)$ \\
$(4,2)$ \\
$(3,1)$ \\
$(2,0)$ \\
.%
\end{tabular}%
\begin{tabular}{l}
$(4,4)$ \\
$(3,3)$ \\
$(2,2)$ \\
$(1,1)$ \\
$(0,0)$%
\end{tabular}%
\begin{tabular}{l}
$(3,5)$ \\
$(2,4)$ \\
$(1,3)$ \\
$(0,2)$ \\
.%
\end{tabular}%
\begin{tabular}{l}
$(2,6)$ \\
$(1,5)$ \\
$(0,4)$ \\
\\
.%
\end{tabular}%
\begin{tabular}{l}
$(1,7)$ \\
$(0,6)$ \\
\\
\\
.%
\end{tabular}%
\begin{tabular}{l}
$(0,8)$ \\
\\
\\
\\
.%
\end{tabular}
\\ \hline\hline
\end{tabular}%
}
\end{equation*}
\end{center}
\end{widetext}

We recall that the raising symplectic generators
$F^{lm}(\alpha,\beta)$ transform according to the $U(6)$ irreducible
representation $[2]_{6}$. According to Eq.(\ref{U6O6}) it decomposes
to the $SO(6)$ irreps $(2)_{6}$ and $(0)_{6}$, respectively.
Further, using Eq.(\ref{O6SUpn3}), we find the $SU_{pn}(3)$ content
for each of these two $SO(6)$ irreps: 1) $(0)_{6}\downarrow (0,0)$;
2)$(2)_{6}\downarrow (2,0), (1,1), (0,2)$. These results actually
coincide with the last two subrows with $\upsilon = 2$ and $0$ of
Table \ref{Ne20-Sp12RIR}. The symplectic lowering generators
$G^{lm}(\alpha,\beta)$ transform according to the conjugate $U(6)$
representation $[2]^{\ast}_{6} = [-2]_{6} \equiv [222220]_{6}$,
which decomposes to the same $SO(6)$ irreps $(2)_{6}$ and $(0)_{6}$.
In turn, we obtain the same $SU_{pn}(3)$ content for the lowering
symplectic generators, as given above for the raising operators. The
product of the lowering and the raising symplectic generators will
transform according to the direct product $[-2]_{6} \otimes [2]_{6}$
of the corresponding $U(6)$ representations, producing the set:
$[2,-2]^{\ast}_{6}$, $[1,-1]^{\ast}_{6}$, and $[0,-0]^{\ast}_{6}$.
Then it is easy to show that acting on the $SU_{pn}(3)$ multiplet
$(8,0)$ by the tensor operator $G^{2}(a,a) \cdot
F^{2}(b,b)=\frac{2}{3}\sqrt{5}[G^{2}(a,a)\times
F^{2}(b,b)]^{4}_{-4100}$, classified by the whole chain
(\ref{O6-DS}), we obtain the $SU_{pn}(3)$ multiplet $(6,2)$. The
operators a's and b's are defined as a linear combination of the
proton and neutron creation or annihilations operators. In
particular, $a^{\dag}_{j}=
\frac{1}{\sqrt{2}}\big(-iB^{\dag}_{j}(p)+B^{\dag}_{j}(n)\big)$ and
$b^{\dag}_{j}=
\frac{1}{\sqrt{2}}\Big(iB^{\dag}_{j}(p)+B^{\dag}_{j}(n)\Big)$
\cite{sp2rxso6} which transform as $(1,0)$ and $(0,1)$ $SU(3)$
tensors, respectively, where the following notation is also used
$B^{\dag}_{i}(\alpha)= \sum_{s}b^{\dagger}_{i\alpha,s}$. Their
conjugate operators $a_{j}$ and $b_{j}$ transform as $(0,1)$ and
$(1,0)$ $SU(3)$ tensors, respectively. By repeated action with the
same operator $G^{2}(a,a) \cdot
F^{2}(b,b)=\frac{2}{3}\sqrt{5}[G^{2}(a,a)\times
F^{2}(b,b)]^{4}_{-4100}$ one can obtain the remaining even
$SU_{pn}(3)$ multiplets belonging to the $SO(6)$ irrep $\upsilon_{0}
= 8$. In this way, the operator $G^{2}(a,a) \cdot
F^{2}(b,b)=\frac{2}{3}\sqrt{5}[G^{2}(a,a)\times
F^{2}(b,b)]^{4}_{-4100}$ can be interpreted as a 2p-2h-like operator
of the core excitations that creates two oscillator quanta in the
shell above and annihilates two oscillator quanta in the shell
bellow, i.e. it promotes two oscillator quanta up. For instance, the
$SU_{pn}(3)$ multiplet $(6,2)$ within the $SO(6)$ irrep
$\upsilon_{0} = 8$ of the symplectic $Sp(12,R)$ bandhead, defined by
$N_{0}$ oscillator quanta, is obtained by promoting two oscillator
quanta from the shell $\mathcal{N}=1$ to $\mathcal{N}=2$, i.e.
changing the many-particle shell-model configuration
$(0)^{4}(1)^{12}(2)^{4}$ to $(0)^{4}(1)^{10}(2)^{6}$, the latter
producing the excited $SU_{pn}(3)$ irrep $(6,2)$ from $(8,0)$ of the
former configuration. The odd $SU_{pn}(3)$ multiplet $(7,1)$ can be
obtained in a similar manner from $(8,0)$ by acting with an
1p-1h-like operator, i.e., by the $U(6)$ operator
$[A^{0}(b,a)]^{2}_{-2100}$. Note that the latter operator together
with $G^{2}(a,a) \cdot
F^{2}(b,b)=\frac{2}{3}\sqrt{5}[G^{2}(a,a)\times
F^{2}(b,b)]^{4}_{-4100}$ both preserve the number of $U(6)$ harmonic
oscillator quanta $N$ of each shell.

\begin{widetext}
\begin{center}
\begin{table}[h!]
\caption{Irreducible collective space 0p-0h $[12]_{6}$ of
$Sp(12,R)$, relevant to $^{106}$Cd, which $SU_{pn}(3)$ basis states
are classified according to the chain (\ref{O6-DS}).}   \label{Cd106-Sp12R-IR} % title of Table
\end{table}
\vspace{-30pt}
%\centering
\smallskip\centering\small\addtolength{\tabcolsep}{-1.pt}
%\begin{tabular}{c c c c}            % centered columns (4 columns)
\begin{equation*}
\smallskip \centering{\small \addtolength{\tabcolsep}{0.7pt}
\begin{tabular}{||l||l||l||}
\hline\hline $\ \ \ N$ &  $\ \ \upsilon \backslash \nu $ &
\begin{tabular}{lllllllllllllllll}
$\cdots $ & $14$ & \ $\ \ \ \ \ 12$ & $\ \ \ \ \ 10$ & $\ \ \ \ \ \
8$ & $\ \ \ \ \ \ \ \ 6 $ & $\ \ \ \ \ \ \ 4$ & $\ \ \ \ \ \ 2$ & $\
\ \ \ \ 0$ & $\ \ \ \ -2$ & \ \ \ \ $-4$ & $\ \ \ \ -6$ & \ \ \ \ \
$-8$ & $\ \ \ \ \ -10$ & $\ \ \ -12$ & $\ \ \ \
-14$ & $\ \ \ \ \cdots $%
\end{tabular}
\\ \hline\hline
$\ \ \ \ \vdots $ & $\ \ \ \ \vdots $ &
\begin{tabular}{lllllllllllllllll}
$\ddots $ & $\ \ \vdots $ & $\ \ \ \ \ \ \ \ \vdots $ & \ \ \ \ \ $\
\ \vdots $ & $\ \ \ \ \ \ \ \ \vdots $ & $\ \ \ \ \ \ \ \ \ \vdots $
& $\ \ \ \ \ \ \ \ \vdots $ & \ \ \ \ \ $\ \vdots $ & $\ \ \ \ \ \
\vdots $ & $\ \ \ \ \ \ \vdots $ & \ \ \ \ \ $\ \ \vdots $ & $\ \ \
\ \ \ \ \vdots $ & $\ \ \ \ \ \ \ \vdots $ & $\ \ \ \ \ \ \ \ \vdots
$ & \ \ \ \ \ \ \ $\ \vdots $ & $\ \ \ \ \ \ \ \ \vdots $ & $\ \ \ \
\
\udots$%
\end{tabular}
\\ \hline
$N_{0}+2$ &
\begin{tabular}{l}
$\ 14$ \\
$\ 12$ \\
$\ 10$ \\
\ $\ \vdots $ \\
$\ \ 2$ \\
\ $\ 0$%
\end{tabular}
&
\begin{tabular}{lllllllllllllllll}
& $(14,0)$ & $(13,1)$ & $(12,2)$ & $(11,3)$ & $(10,4) $ & $(9,5)$ &
$(8,6)$ & $(7,7)$ & $(6,8)$ & $(5,9)$ & $(4,10) $ & $(3,11)$ &
$(2,12)$ & $(1,13)$ &
$(0,14)$ &  \\
&  & $(12,0)$ & $(11,1)$ & $(10,2)$ & $(9,3) $ & $(8,4)$ & $(7,5)$ &
$(6,6)$
& $(5,7)$ & $(4,8)$ & $(3,9) $ & $(2,10)$ & $(1,11)$ & $(0,12)$ &  &  \\
&  &  & $(10,0)$ & $\ (9,1)$ & $(8,2) $ & $(7,3)$ & $(6,4)$ & $(5,5)$ & $%
(4,6)$ & $(3,7)$ & $(2,8) $ & $(1,9)$ & $(0,10)$ &  &  &  \\
&  &  &  & $\ \ \ \ \ \ \ \ \ddots $ &  &  & $\ \ \ \vdots $ & $\ \
\ \vdots $ & $\ \ \ \vdots $ &  &  & $\udots$ &  &  &  &  \\
&  &  &  &  &  &  & $(2,0)$ & $(1,1)$ & $(0,2)$ &  &  &  &  & & &
\\
&  &  &  &  &  &  &  & $(0,0)$ &  &  &  &  &  &  &  &
\end{tabular}
\\ \hline
$\ \ N_{0}$ &
\begin{tabular}{l}
\ $12$ \\
$\ 10$ \\
$\ \ \vdots $ \\
$\ \ 2$ \\
\ $\ 0$%
\end{tabular}
& \ \ \ \ \ \ \ \ \ \ \ \
\begin{tabular}{lllllllllllll}
$(12,0)$ & $(11,1)$ & $(10,2)$ & $(9,3) $ & $(8,4)$ & $(7,5)$ & $(6,6)$ & $%
(5,7)$ & $(4,8)$ & $(3,9) $ & $(2,10)$ & $(1,11)$ & $(0,12)$ \\
& $(10,0)$ & $\ (9,1)$ & $(8,2) $ & $(7,3)$ & $(6,4)$ & $(5,5)$ &
$(4,6)$ &
$(3,7)$ & $(2,8) $ & $(1,9)$ & $(0,10)$ &  \\
&  & $\ \ \ \ \ \ \ \ \ddots $ &  &  & $\ \ \ \vdots $ & $\ \ \
\vdots $ & $\ \ \ \vdots $ &  &  & $\udots$ &  &  \\
&  &  &  &  & $(2,0)$ & $(1,1)$ & $\ (0,2)$ &  &  &  &  &  \\
&  &  &  &  &  & $(0,0)$ &  &  &  &  &  &
\end{tabular}
\\ \hline\hline
\end{tabular}%
}
\end{equation*}
\end{center}
\end{widetext}

Thus, we have seen that the $SU(3)$ many-particle shell-model
configurations are organized in different way by means of the group
$SO(6)$ through the reduction $U(6) \supset SO(6) \supset SU(3)$
(more precisely, $Sp(12,R) \supset U(6) \supset SO(6) \supset
SU_{pn}(3) \otimes SO(2)$ for different $U(6)$ shells), compared to
the standard shell-model plethysm reduction $U(d) \supset SU(3)$.
Each horizontal subset of the $SU(3)$ multiplets is now
characterized by the same value of the $SO(6)$ irrep
$\upsilon=\lambda+\mu$. In this regard, we want to point out that
the $SU(3)$ content of the $U(6)$ shells defined by the PNSM
dynamical chain $Sp(12,R) \supset U(6) \supset SU_{p}(3)\otimes
SU_{n}(3) \supset SU(3)$ considered, e.g., in Refs.\cite{cdf,smpnsm}
will coincide precisely with that generated first by the separate
reductions $U_{\alpha}(d) \supset SU_{\alpha}(3)$ ($\alpha=p,n$)
with the subsequent coupling of the proton $(\lambda_{p},\mu_{p})$
and neutron $(\lambda_{n},\mu_{n})$ subsystem representations to the
combined proton-neutron $SU(3)$ irreducible representation
$(\lambda,\mu)$ since in this case the PNSM many-particle $SU(3)$
configurations are organized by means of the group structure
$SU_{p}(3)\otimes SU_{n}(3) \supset SU(3)$ within the $U(6)$
harmonic oscillator shell. For $^{20}$Ne, the $U(6)$ irrep $[8]_{6}$
according to the underlying algebraic structure $SU_{p}(3)\otimes
SU_{n}(3) \supset SU(3)$ produces the three sets: 1) $(4,0)\otimes
(4,0) = (8,0), (6,1), (4,2), (2,3), (0,4)$; 2) $(4,0)\otimes (0,2) =
(4,2), (3,1), (2,0)$; 3) $(0,2)\otimes (0,2) = (0,4), (1,2), (2,0)$.
The latter are exactly those obtained by means of the plethysm
operation given earlier.

The $SU_{pn}(3)$ states belonging to the $SO(6)$ representations
with $\upsilon > \upsilon_{0}$ within the irreducible collective
space of $Sp(12,R)$ can be obtained from those belonging to the
$SO(6)$ irrep $\upsilon_{0}$ by acting with the raising symplectic
generators $F^{lm}(\delta,\tau)$ ($\delta,\tau = a,b$). In
particular, the $SU_{pn}(3)$ states with $\upsilon=\upsilon_{0}+2$
belonging to the next $U(6)$ shell are readily obtained by acting on
the $SU_{pn}(3)$ states contained in the $SO(6)$ representation
$\upsilon_{0}$ with the raising generators $F^{2m}(a,a)$,
$F^{2m}(a,b)$, and $F^{2m}(b,b)$ which transform as $(2,0)$, $(1,1)$
and $(0,2)$ $SU(3)$ tensors, respectively. The construction of the
remaining $SU(3)$ basis states of the irreducible collective space
of $Sp(12,R)$ then becomes straightforward.

In this way, we have shown that the $SU_{pn}(3)$ states within the
$SO(6)$ irrep $\upsilon_{0}$ are Pauli allowed. Then the states
generated from them and belonging to $\upsilon > \upsilon_{0}$ are
also Pauli allowed. The states with $\upsilon < \upsilon_{0}$ are
not particularly interesting, but we will make a comment concerning
them. First, note that each of the $SU(3)$ multiplets in the $U(d)
\supset SU(3)$ set $(8,0), (4,2), (0,4), (2,0)$ is contained
respectively in the $SO(6)$ representation with $\upsilon = 8, 6,
4,$ and $2$ of the $U(6)$ irrep $[8]_{6}$, given in Table
\ref{Ne20-Sp12RIR} by the row with $N_{0}$. The state $(0,0)$, for
this specific example of $^{20}$Ne, is missing in the $U(d) \supset
SU(3)$ set. Thus it is Pauli forbidden. The $SU_{pn}(3)$ states,
which are placed to the right of the Pauli allowed multiplets
$(8,0), (4,2), (2,0)$ at each row that is defined by the
corresponding $SO(6)$ irrep with $\upsilon < \upsilon_{0}$ are Pauli
allowed, since they also can be represented as
multi-particle-multi-hole excitations built upon the $(8,0)$,
$(4,2)$, or $(2,0)$, correspondingly. The $SU_{pn}(3)$ states on the
left from these multiplets are Pauli forbidden, since they
correspond to promoting two oscillator quanta down to filled shells.

Up to now, nothing was said about the spin content. In this respect,
we remind that the proper permutational symmetry in the PNSM is
ensured by the reduction
\begin{equation}
O(m) \supset S_{A} \label{Sn}
\end{equation}
of the complementary group $O(m)$ in the reduction of the
many-particle dynamical group $Sp(12m,R)$ of the whole system, i.e.,
$Sp(12m,R) \supset Sp(12,R) \otimes O(m)$ \cite{cdf,smpnsm}. The
irrep $\omega$ of $O(m)$ is determined by the irrep $\langle\sigma
\rangle$ of $Sp(12,R)$ and vice versa. But, since the antisymmetry
should be satisfied separately for protons and neutrons, in order to
insure the proper permutational symmetry, we consider further the
reduction \cite{cdf,smpnsm}:
\begin{align}
&O(m) \quad \supset \quad S_{A} \quad \supset \quad S_{Z} \quad
\otimes \quad S_{N}, \label{SZxSN}\\
&\quad\omega \qquad \delta \quad \ [f] \quad \delta_{0} \quad
[f_{p}] \qquad \ \ [f_{n}] \notag
\end{align}
where the quantum numbers bellow different groups stand for their
irreducible representations, $\delta$ and $\delta_{0}$ are
multiplicity indices. Due to the overall antisymmetry, the spin wave
functions for the proton and neutron subsystems are characterized by
the conjugate representations $[\widetilde{f}_{p}]$ and
$[\widetilde{f}_{n}]$ \cite{cdf,smpnsm}, respectively. In the
standard shell model, using the proton-neutron formalism, the spin
content is obtained by considering the reduction $U_{\alpha}(2d)
\supset U_{\alpha}(d) \otimes U_{S_{\alpha}}(2)$ ($\alpha = p,n$).
For example, in the case of $^{20}Ne$, the maximal spatial symmetry
of two protons in the $sd$ shell is provided by the $U_{p}(6)$ irrep
$[f_{p}]=[2]$, which in addition must be compatible with the
permutational symmetry of the whole proton subsystem, i.e.
$[f_{Z}]=[f_{0}][f_{1}][f_{2}]\equiv [2^{5}]$ with $f_{i}$ related
to a single-shell configuration $(i)^{p_{i}}$. Then, the conjugate
proton spin symmetry is $[\widetilde{f}_{p}] = [11]$, from which one
obtains $S_{p} =\frac{1}{2}(\widetilde{f}_{1p}-\widetilde{f}_{2p})=
0$. Similar considerations are valid for the neutron subsystem,
i.e., $S_{n} = 0$. Hence the total spin is also zero. In the PNSM,
the $O(19)$ irrep of  $^{20}Ne$ is determined by the $Sp(12,R)$
bandhead $\sigma = [10,2,2,2,2,2]_{6}$, i.e. $\omega =
(10,2,2,2,2,2)$. Using the computer program \cite{SCHUR} and the
Pauli allowed spatial symmetries of the type $[444\ldots422\ldots2]$
for even-even nuclei, one sees that the $O(19)$ irrep
$(10,2,2,2,2,2)$ reduces to the maximal space symmetry $S_{20}$
irrep [444422], which in turn reduces to the $S_{10} \otimes S_{10}$
irrep $[2^{5}][2^{5}]$. Then from the proton spatial symmetry
$[f_{p}]=[2^{5}]$ one obtains the conjugate spin symmetry
$[\widetilde{f}_{p}]=[55]$, from which it follows that $S_{p}=0$.
Similarly, one obtains $S_{n}=0$, and hence, $S=0$. In this way,
similarly to the standard shell model, within the PNSM the spatial
symmetry is also accompanied by the corresponding conjugate spin
symmetry. We will not consider further the spin content, which can
be recovered when this is required. We note too that because the
$Sp(12,R)$ generators are $O(m)$-scalar operators, they are also
$S_{A}$-scalar operators, and therefore they preserve the
permutational symmetry.

We recall that to account for the four Pauli allowed $SU(3)$
multiplets $(8,0), (4,2), (0,2), (2,0)$  of the shell
$\mathcal{N}=2$, one needs to consider the direct sum of these four
irreducible collective many-particle subspaces in the Elliott
$SU(3)$ or $Sp(6,R)$ shell models. In the microscopic shell-model
version of the BM model they result in a single $Sp(12,R)$ irrep.
The Pauli allowed $SU_{pn}(3)$ states with $\upsilon \geq
\upsilon_{0}$ of the $Sp(12,R)$ irreducible collective space 0p-0h
$[8]_{6}$ were used in Ref. \cite{Ne20} for the description of the
ground and first two beta bands in $^{20}$Ne by using both vertical
and horizontal mixing interaction.

\subsection{Shell-model representations of $^{106}$Cd}

As a second example, we consider the weakly deformed nucleus
$^{106}$Cd. For heavy nuclei we use the pseudo-$SU(3)$ scheme
\cite{pseudoSU3a,pseudoSU3b,pseudoSU3c}. Filling pairwise the
pseudo-Nilsson levels with protons at observed quadrupole
deformation $\beta \approx 0.17$ \cite{Raman01} one obtains
completely filled $\mathcal{\widetilde{N}}=2$ pseudo-shell plus 8
protons in the unique-parity level $g_{9/2}$. Then the leading
proton $SU_{p}(3)$ irrep is the scalar irrep $(0,0)$, since the
particles in the unique-parity levels in the pseudo-$SU(3)$ scheme
are considered in seniority zero configuration. Similarly, for
neutrons one obtains completely filled $\mathcal{\widetilde{N}}=2$
pseudo-shell plus 6 (or 8) neutrons occupying the
$\mathcal{\widetilde{N}}=3$ pseudo-shell and 2 (or 0) neutrons in
the unique-parity level $h_{11/2}$. Using the codes
\cite{UNtoSU3a,UNtoSU3b}, one obtains the set of Pauli allowed
$SU(3)$ states: $(12,0), (9,3), (6,6), (7,4), (8,2), \ldots$ or
$(10,4), (12,0), (8,5), (9,3), (10,1), (5,8), (6,6), (7,4)$, $(8,2),
\ldots$ considering 6 or 8 active neutrons, respectively. Further,
the proton and neutron irreps should be coupled to obtain the
combined proton-neutron $SU_{pn}(3)$ representation of the whole
nuclear system. But since for the proton subsystem only the scalar
representation $(0,0)$ is admitted, the set of combined
proton-neutron multiplets coincides with that of the neutron
subsystem since $(\lambda_{p},\mu_{p})\otimes (\lambda_{n},\mu_{n})=
(0,0)\otimes (\lambda_{n},\mu_{n}) \equiv (\lambda,\mu)$.

Alternatively, using the supermultiplet scheme one readily obtains
that 6 nucleons fill the last valence $\mathcal{\widetilde{N}}=3$
pseudo-shell. Again using the codes \cite{UNtoSU3a,UNtoSU3b}, one
gets the set of $SU(3)$ states:
\begin{align}
&(14,2), (12,3), (13,1), (10,4), (11,2), (12,0), \ldots,
\notag\\
&(9,3), (6,6), \ldots, (8,2), (5,5), (2,8), \ldots, \notag\\
&(7,1), (4,4), (1,7), \ldots, (6,0), (3,3), (0,6), \ldots, \notag\\
&(2,2), (1,1), (0,0), \ldots \label{nucleons-SUIRs}
\end{align}
Using the Nilsson model ideas
\cite{Carvalho92,StrApprox1,Park84,Jarrio91}, we choose the $SU(3)$
irrep $(12,0)$, contained in the two alternatively obtained sets of
Pauli allowed many-particle $SU(3)$ states, by means of which we fix
the appropriate $Sp(12,R)$ irreducible representation 0p-0h
$[12]_{6}$ (or $\langle \sigma\rangle = \langle 27+m/2, 15+m/2,
\ldots, 15+m/2\rangle$ using an equivalent notation) which turns to
be useful for the description of the low-energy quadrupole
collectivity observed in $^{106}$Cd. The relevant irreducible
collective space for $^{106}$Cd, spanned by the $Sp(12,R)$
irreducible representation 0p-0h $[12]_{6}$ restricted to the fully
symmetric $U(6)$ irreps only and which $SU_{pn}(3)$ basis states are
classified according to the chain (\ref{O6-DS}), is given in Table
\ref{Cd106-Sp12R-IR}. By comparing the set (\ref{nucleons-SUIRs})
with the $SU_{pn}(3)$ multiplets of the symplectic bandhead
structure for the $SO(6)$ irreps $\upsilon < \upsilon_{0}=12$ given
in Table \ref{Cd106-Sp12R-IR} at the row $N_{0}$, one now sees, in
contrast to the case of $^{20}$Ne, that there are many more Pauli
allowed states, including the scalar $SU_{pn}(3)$ irrep $(0,0)$.
Recall, that all $SU_{pn}(3)$ multiplets that are on the right from
a giving $(\lambda',\mu')$ multiplet at each $\upsilon <
\upsilon_{0}$, which is matched with a certain $SU(3)$ irrep of the
set (\ref{nucleons-SUIRs}), are Pauli allowed. In practical
applications, however, only the $SU_{pn}(3)$ states with $\upsilon
\geq \upsilon_{0}$ are used in the diagonalization of the model
Hamiltonian. Hence, only the Pauli allowed states are retained in
the many-particle irreducible collective space of a given $Sp(12,R)$
shell-model representation.

\begin{widetext}
\begin{center}
\begin{table}[h!]
\caption{Irreducible collective space 0p-0h $[36]_{6}$ of
$Sp(12,R)$, relevant to $^{158}$Gd, which $SU_{pn}(3)$ basis states
are classified according to the chain (\ref{O6-DS}).}   \label{Gd158-Sp12RIR} % title of Table
\end{table}
\vspace{-30pt}
%\centering
\smallskip\centering\small\addtolength{\tabcolsep}{1.pt}
%\begin{tabular}{c c c c}            % centered columns (4 columns)
\begin{equation*}
\smallskip \centering{\small \addtolength{\tabcolsep}{-0.8pt}
\begin{tabular}{||l||l||l||}
\hline\hline $\ \ \ \ N$ &  $\ \upsilon \backslash \nu $ &
\begin{tabular}{lllllllllllllllll}
$\cdots $ &  $\ \ 38$ &  \ $\ \ \ 36$ &  $\ \ \ \ \ 34$ &  $\ \ \ \
\ \ 32$ &  $\ \ \cdots $ &  $\ \ \ \ \ 4$ &  $\ \ \ \ \ \ \ 2$ &  $\
\ \ \ \ \ \ \ \
0$ &  $\ \ \ \ \ \ \ -2$ &  \  \ \ \ \ \ \  $-4$ &  $\ \ \ \ \cdots $ &  \  $%
-32$ &  $\ \ \ \ -34$ &  $\ \ \ -36$ &  $\ \ \ \ -38$ &  $\ \cdots $%
\end{tabular}
\\ \hline\hline
$\ \ \ \ \vdots $ &  $\ \ \vdots $ &
\begin{tabular}{lllllllllllllllll}
$\ddots $ &  $\ \ \ \ \vdots $ &  $\ \ \ \ \ \ \vdots $ &  \ \ \ \
$\ \ \ \vdots $ &  $\ \ \ \ \ \ \ \ \vdots $ &  $\ \ \ \ \cdots $ &
$\ \ \ \ \ \vdots $ &  \ \ \ \ \ \ \  $\ \vdots $ &  $\ \ \ \ \ \ \
\ \ \vdots $ &  $\ \ \ \ \ \ \ \ \ \ \ \vdots $ &  \ \ \ \ \ \ \ \ \
$\ \vdots $ &  $\ \ \
\cdots $ &  $\ \ \ \ \ \vdots $ &  $\ \ \ \ \ \ \ \vdots $ &  \ \ \ \ \ \  $%
\ \vdots $ &  $\ \ \ \ \ \ \ \ \vdots $ &  $\ \ \ \udots $%
\end{tabular}
\\ \hline
$N_{0}+2$ &
\begin{tabular}{l}
$\ 38$ \\
$\ 36$ \\
$\ 34$ \\
\  $\ \vdots $ \\
$\ 2$ \\
\  $0$%
\end{tabular}
&
\begin{tabular}{lllllllllllllllll}
& $(38,0)$ & $(37,1)$ & $(36,2)$ & $(35,3)$ & $\cdots $ & $(21,17)$ & $%
(20,18)$ & $(19,19)$ & $(18,20)$ & $(17,21)$ & $\cdots $ & $(3,35)$
& $(2,36)
$ & $(1,37)$ & $(0,38)$ &  \\
&  & $(36,0)$ & $(35,1)$ & $(34,2)$ & $\cdots $ & $(20,16)$ & $(19,17)$ & $%
(18,18)$ & $(17,19)$ & $(16,20)$ & $\cdots $ & $(2,34)$ & $(1,35)$ &
$(0,36)$
&  &  \\
&  &  & $(34,0)$ & $(33,1)$ & $\cdots $ & $(19,15)$ & $(18,16)$ &
$(17,17)$
& $(16,18)$ & $(15,19)$ & $\cdots $ & $(1,33)$ & $(0,34)$ &  &  &  \\
&  &  &  &  $\ \ \ \ddots $ &  &  &  $\ \ \ \ \ \vdots $ &  $\ \ \ \
\vdots $
&  $\ \ \ \ \ \vdots $ &  &  &  $\ \ \udots $ &  &  &  &  \\
&  &  &  &  &  &  &   $\ (2,0)$ &  $\ (1,1)$ &  $\ (0,2)$ &  &  &  &
&  &
&  \\
&  &  &  &  &  &  &  &  $\ (0,0)$ &  &  &  &  &  &  &  &
\end{tabular}
\\ \hline
$\ \ N_{0}$ &
\begin{tabular}{l}
\  $36$ \\
$\ 34$ \\
$\ \ \vdots $ \\
$\ \ 2$ \\
\ $\ 0$%
\end{tabular}
&  \ \ \ \ \ \ \ \ \ \ \
\begin{tabular}{lllllllllllll}
$(36,0)$ & $(35,1)$ & $(34,2)$ & $\cdots $ & $(20,16)$ & $(19,17)$ &
$(18,18)
$ & $(17,19)$ & $(16,20)$ & $\cdots $ & $(2,34)$ & $(1,35)$ & $(0,36)$ \\
& $(34,0)$ & $(33,1)$ & $\cdots $ & $(19,15)$ & $(18,16)$ & $(17,17)$ & $%
(16,18)$ & $(15,19)$ & $\cdots $ & $(1,33)$ & $(0,34)$ &  \\
&  &  $\ \ \ \ddots $ &  &  &  $\ \ \ \ \ \vdots $ &  $\ \ \ \ \
\vdots $ &
$\ \ \ \ \ \vdots $ &  &  &  $\ \ \udots $ &  &  \\
&  &  &  &  &   $\ (2,0)$ &  $\ (1,1)$ &  $\ (0,2)$ &  &  &  &  &  \\
&  &  &  &  &  &  $\ (0,0)$ &  &  &  &  &  &
\end{tabular}
\\ \hline\hline
\end{tabular}%
}
\end{equation*}
\end{center}
\end{widetext}

\subsection{Shell-model representations of $^{158}$Gd}

As a final example, we consider the strongly deformed nucleus
$^{158}$Gd. Similarly, using the pseudo-$SU(3)$ scheme
\cite{pseudoSU3a,pseudoSU3b,pseudoSU3c} one obtains the following
many-particle configurations: 1)
$(\widetilde{2})^{20}(\widetilde{3})^{8}$ plus 6 protons occupying
the unique-parity level $h_{11/2}$; and 2)
$(\widetilde{2})^{20}(\widetilde{3})^{20}(\widetilde{4})^{6}$ plus 6
neutrons occupying the unique-parity level $i_{13/2}$. The codes
\cite{UNtoSU3a,UNtoSU3b} produce the following two sets: a) $(10,4),
(12,0), (8,5), (9,3), \dots$; and b) $(18,0), (15,3), (12,6),
(13,4), (14,2), \ldots$. Coupling the leading $SU_{p}(3)$ and
$SU_{n}(3)$ irreps, i.e., $(10,4) \otimes (18,0)$, one obtains the
leading (most deformed) combined proton-neutron multiplet $(28,4)$.
Alternatively, using the many-particle configuration
$(\widetilde{2})^{40}(\widetilde{3})^{34}$ (plus 24 nucleons
occupying unique-parity level $h_{11/2}$) based on the
pseudo-$SU(3)$ and supermultiplet schemes, one obtains the following
set of $SU(3)$ states: $(2,14), (3,12), (4,10), (5,8), (6,6), (7,4),
(8,2), (1,13)$, $(2,11), (3,9), \ldots$, which consists of
predominantly oblate-like $SU(3)$ multiplets with $\lambda < \mu$.
The most deformed prolate-like $SU(3)$ irreducible representations
are $(7,4)$ and $(8,2)$. The leading $SU(3)$ multiplet is $(2,14)$,
corresponding to oblate-like shape of the combined proton-neutron
nuclear system. Hence the supermultiplet scheme of filling the
pseudo-Nilsson levels at the observed quadrupole deformation is not
appropriate for this nucleus. This is a well known result for
nuclei, in which the valence protons and neutrons occupy different
shells. Using again the Nilsson model ideas
\cite{Carvalho92,StrApprox1,Park84,Jarrio91}, we choose the
axially-symmetric prolate $SU(3)$ representation $(36,0)$, which is
now a linear combination of Slater determinants. The latter is
obtained by coupling axially-symmetric proton and neutron
multiplets, i.e.,
\begin{align}
(18,0) \otimes (18,0) = & \ (36,0), (34,1), (32,2), (30,3), \notag\\
& \ \ldots, (2,17), (0,18). \label{nucleons-SUIRs-Gd158}
\end{align}
The relevant irreducible collective space for $^{158}$Gd, spanned by
the $Sp(12,R)$ irreducible representation 0p-0h $[36]_{6}$
restricted to the fully symmetric $U(6)$ irreps only and which
$SU_{pn}(3)$ basis states are classified according to the chain
(\ref{O6-DS}), is given in Table \ref{Gd158-Sp12RIR}. Similar
considerations are valid also, concerning the $SU_{pn}(3)$
multiplets belonging to the $SO(6)$ irreps with $\upsilon <
\upsilon_{0}$ for $^{158}$Gd, not all of which are Pauli permitted.
For instance, those which are on the left from the $SU_{pn}(3)$
multiplets $(\lambda',\mu')$, the latter matching the corresponding
irreps from the set (\ref{nucleons-SUIRs-Gd158}), are not allowed
for $36 < \upsilon \leq 18$, as well as all $SU_{pn}(3)$ multiplets
belonging to $\upsilon \leq 16$. The Pauli allowed $SU_{pn}(3)$
multiplets contained in the maximal seniority $SO(6)$ irrep
$\upsilon_{0}=36$ of the symplectic bandhead were used for studying
the low-energy quadrupole dynamics in $^{158}$Gd \cite{Gd158} by
using a horizontal mixing interaction in the model Hamiltonian.

\section{Conclusions}

The structure of the irreducible collective space of the $Sp(12,R)$
shell-model representations with respect to the $SU(3)$ symmetry of
the many-particle nuclear states, which are classified by the
dynamical chain $Sp(12,R) \supset U(6) \supset SO(6) \supset
SU_{pn}(3) \otimes SO(2) \supset SO(3)$ within the framework of the
PNSM, is considered in detail. This chain has been shown to
correspond to a microscopic shell-model version of the BM model,
obtained recently by embedding the original BM model into the
two-component proton-neutron shell-model theory. The construction of
the relevant shell-model representations of the $Sp(12,R)$ dynamical
group has been considered for three nuclei with varying collective
properties belonging to different mass regions. A comparison with
the standard consideration of the Pauli allowed $SU(3)$ states
within the valence shells using the proton-neutron formalism, or
within a single valence shell using the supermultiplet scheme, both
exploiting the plethysm operation defined by the reduction $U(d)
\supset SU(3)$ with $d =\frac{1}{2}(\mathcal{N}+1)(\mathcal{N}+2)$
for any major shell $\mathcal{N}$, is given. It was shown that, in
the present proton-neutron shell-model approach, the $SU_{pn}(3)$
many-particle states are organized in a different way into different
$SO(6)$ subshells of the six-dimensional harmonic oscillator with a
given $SO(6)$ irreducible representation $\upsilon=\lambda+\mu$.
This is in contrast to the case of filling the levels of the
standard three-dimensional harmonic oscillator and using the
plethysm operation. In this way, in contrast to the standard
proton-neutron shell-model reduction $U_{\alpha}(2d) \supset
[U_{\alpha}(d) \supset SU_{\alpha}(3)] \otimes [U_{S_{\alpha}}(2)
\supset SU_{S_{\alpha}}(2)]$, new structures and organization
(coupling schemes) of the many-particle nuclear states appear within
the framework of the PNSM with the direct-product dynamical group
$Sp(12,R) \otimes O(m) \subset Sp(12m,R)$. Different subgroups of
$Sp(12,R)$ enrich the dynamical content of possible collective
motions in the two-component proton-neutron nuclear systems, whereas
the $O(m)$ group ensures the proper permutational symmetry.

Further, it was shown that the $SU_{pn}(3)$ states belonging to the
$SO(6)$ irreps $\upsilon \geq \upsilon_{0}$ are always Pauli
permitted, whereas for $\upsilon < \upsilon_{0}$ not all of the
$SU_{pn}(3)$ multiplets are Pauli allowed. $\upsilon_{0}$ denotes
the maximal seniority $SO(6)$ irreducible representation contained
in the symplectic $Sp(12,R)$ bandhead. The situation, in some
respect, resembles that encountered in the algebraic cluster models
(see, e.g., Refs. \cite{SACMa,SACMb}) based on the $SU(3)$ dynamical
group, for which the Pauli allowed $SU(3)$ states are restricted
from below by the minimal number of oscillator quanta $n_{0}$. All
states with $n < n_{0}$ are Pauli forbidden, and hence they are
excluded from the many-particle subspaces of the Hilbert space for
the cluster configuration under consideration. For the PNSM which
states are classified by the chain (\ref{O6-DS}), we also have a
restriction of the many-particle states by the minimal Pauli allowed
number of harmonic oscillator quanta $N_{0}$. But due to the
repeating substructure for each subsequent $U(6)$ representation
(cf. Tables \ref{Ne20-Sp12RIR}$-$\ref{Gd158-Sp12RIR}), resulting
from the properties of the group $SO(6)$ in the reduction $U(6)
\supset SO(6) \supset SU(3)$, the restriction with respect to
$N_{0}$ is not enough. One needs similar restriction with respect to
the $SO(6)$ quantum number, although not all of the corresponding
$SU_{pn}(3)$ states belonging to the $SO(6)$ irreps $\upsilon <
\upsilon_{0}$ are Pauli forbidden. For any case, it is safe to
discard them and to consider only the Pauli allowed $SU_{pn}(3)$
multiplets with $\upsilon \geq \upsilon_{0}$ spanning the
many-particle irreducible collective space of the relevant
$Sp(12,R)$ shell-model representation of the concrete nuclear
system. The latter are exactly those $SU_{pn}(3)$ many-particle
states that are exploited in the practical applications of the
microscopic shell-model version of the BM model.

Finally, it should be pointed out that for the medium-mass and heavy
nuclei, one may alternatively use the proxy-$SU(3)$ scheme
\cite{proxySU3} (see also, e.g., Ref. \cite{Bonatsos23} for a recent
review) instead of the pseudo-$SU(3)$ one, as it was done, for
instance, for the description of the irrotational-flow quadrupole
dynamics in $^{102}$Pd in Ref. \cite{Pd102}.


\begin{thebibliography}{9}

\bibitem{Elliott58} J. P. Elliott,
Proc. R. Soc. \textbf{A 245}, 128 (1958); \textbf{245}, 562 (1958).

\bibitem{BM} A. Bohr and B. R. Mottelson, \emph{Nuclear Structure} (W.A. Benjamin
Inc., New York, 1975), Vol. II.

\bibitem{Bohr52} A. Bohr, Mat. Fys. Medd. Dan. Vid. Selsk. \textbf{26} (14)
(1952).

\bibitem{BM53} A. Bohr and B. R. Mottelson, Mat. Fys. Medd. Dan. Vid. Selsk. \textbf{27} (16)
(1953).

\bibitem{IBM} F. Iachello and A. Arima, \textit{The Interacting Boson Model}
(Cambridge University Press, Cambridge, 1987).

\bibitem{stretched} D. J. Rowe, Rep. Prog. Phys. \textbf{48}, 1419
(1985).

\bibitem{Rowe96} D. J. Rowe, Prog. Part. Nucl. Phys. \textbf{37},
265 (1996).

\bibitem{DGSGA} \emph{Dynamical Groups and Spectrum Generating Algebras} (in 2 Volumes)
by A. Bohm, Y. Ne'eman, A.O. Barut and others, (World Scientific,
Singapore, 1988).

\bibitem{Heyde94} K. L.G. Heyde, \textit{The Nuclear Shell Model}
(Springer-Verlag, Berlin Heidelberg, 1994).

\bibitem{cm3} L. Weaver, L. C. Biedenharn, and R. Y. Cusson, Ann.
Phys. (N.Y.) \textbf{77}, 250 (1973).

\bibitem{NA4} O. L. Weaver, R. Y. Cusson and L. C. Biedenharn, Ann. Phys. (N.Y.) \textbf{102},
493 (1976).

\bibitem{RR1} G. Rosensteel and D. J. Rowe,
Phys. Rev. Lett. \textbf{38}, 10  (1977).

\bibitem{IVBM} A. Georgieva, P. Raychev, and R. Roussev,
J. Phys. \textbf{G8}, 1377 (1982).

\bibitem{cdf} H. G. Ganev, Eur. Phys. J. \textbf{A 50}, 183 (2014).

\bibitem{smpnsm} H. G. Ganev, Eur. Phys. J. A \textbf{51}, 84 (2015).

\bibitem{microBM} H. G. Ganev, Eur. Phys. J. \textbf{A 57}, 181 (2021).

\bibitem{mbm-cr} H. G. Ganev, Chin. Phys. \textbf{C 47}, 104101  (2023).

\bibitem{MQ70}  M. Moshinsky and C. Quesne,
J. Math. Phys. \textbf{11}, 1631 (1970).

\bibitem{sp2rxso6} H. G. Ganev, Chin. Phys. \textbf{C 45}, 114101  (2021).

\bibitem{Van71} V. V. Vanagas,
\emph{Algebraic methods in nuclear theory} (Mintis, Vilnius, 1971)
(in Russian).

\bibitem{SCHUR} B. G. Wybourne, SCHUR, An interactive program for calculating
properties of Lie groups and symmetric functions, Euromath Bulletin
\textbf{2}, 145 (1996); http://schur.sourceforge.net/

\bibitem{UNtoSU3a} J. P. Draayer, Y. Leschber, S. C. Park, R. Lopez, Comput. Phys.
Commun. \textbf{56}, 279 (1989).

\bibitem{UNtoSU3b} D. Langr, T. Dytrych, J. P. Draayer, K. D.
Launey, and P. Tvrdik, Comput. Phys. Commun. \textbf{244}, 442
(2019).

\bibitem{Ne20} H. G. Ganev, Chin. Phys. \textbf{C 46}, 044105 (2022).

\bibitem{pseudoSU3a} R. D. Ratna Raju, J. P. Draayer, and K. T. Hecht,
Nucl. Phys. \textbf{A 202}, 433 (1973).

\bibitem{pseudoSU3b} J. P. Draayer and K. J. Weeks,
Phys. Rev. Lett. \textbf{51}, 1422 (1983).

\bibitem{pseudoSU3c} J. P. Draayer and K. J. Weeks,
Ann. Phys. \textbf{156}, 41 (1984).

\bibitem{Raman01} S. Raman, C. W. Nestor, Jr, and P. Tikkanen, Atomic Data and Nuclear
Data Tables \textbf{78}, 1 (2001).

\bibitem{Carvalho92} J. Carvalho and D. J. Rowe,
Nucl. Phys. \textbf{A 548}, 1 (1992).

\bibitem{StrApprox1} J. Carvalho, P. Park, D. J.
Rowe, and G. Rosensteel, Phys. Lett. \textbf{B 119}, 249 (1982).

\bibitem{Park84} P. Park, J. Carvalho, M. Vassanji, D. J. Rowe, and
G. Rosensteel, Nucl. Phys. \textbf{A 414}, 93 (1984).

\bibitem{Jarrio91} M. Jarrio, J. L. Wood, and D. J. Rowe,
Nucl. Phys. \textbf{A 528}, 409 (1991).

\bibitem{Gd158} H. G. Ganev, Int. J. Mod. Phys.
\textbf{E 31},  2250047 (2022).

\bibitem{SACMa} J. Cseh, Phys. Lett. \textbf{B 281}, 173 (1992).

\bibitem{SACMb} J. Cseh and G. Levai, Ann. Phys. (N.Y.) \textbf{230}, 165
(1994).

\bibitem{proxySU3} D. Bonatsos, I. E. Assimakis, N.
Minkov, A. Martinou, R. B. Cakirli, R. F. Casten, and K. Blaum,
Phys. Rev. \textbf{C 95}, 064325 (2017).

\bibitem{Bonatsos23} D. Bonatsos et al., Symmetry
\textbf{15}, 169 (2023).

\bibitem{Pd102} H. G. Ganev, Chin. Phys. \textbf{C
48}, 014102 (2024).


\end{thebibliography}
\end{document}